\documentclass[aps,showkeys,superscriptaddress]{revtex4}

\begin{document}
\title{ $CP^{1|1}$ nonlinear sigma model vs strong electron correlations}
\author{E.A. Kochetov}
\affiliation{International Institute of Physics - UFRN, Natal, Brazil;\\
Laboratory of Theoretical Physics, Joint
Institute for Nuclear Research, 141980 Dubna, Russia}
\author{A. Ferraz}
\affiliation{International Institute of Physics - UFRN,
Department of Experimental and Theoretical Physics - UFRN, Natal, Brazil}

\begin{abstract}
The nonlinear sigma model targeted on the coset
supermanifold $CP^{1|1}=SU(2|1)/U(1|1)$
is derived in an attempt to describe the quasiclassical low-energy effective action for the
doped $t-J$ model at the SUSY point, $J=2t.$
In spite of the fact that the supermanifold $CP^{1|1}$ indeed appears as the phase space of the strongly
correlated electrons, the canonical $CP^{1|1}$ nonlinear sigma model (NLSM) is unable to capture
the physics of strong correlations displayed by the SUSY $t-J$  model at any finite doping.
This is due to the fact that, in this regime, the doping itself
cannot be self-consistently incorporated into the $CP^{1|1}$ NLSM.
\end{abstract}

\keywords{$t-J$ model of strongly correlated electrons, $su(2|1)$ coherent state,
$CP^{1|1}$ nonlinear sigma model}
\maketitle

\section{Introduction: $t-J$ model of strongly correlated electrons}

As is well known, a $d$ - dimensional quantum antiferromagnetic (AF)
Heisenberg model in the large-spin  limit at zero temperature
can be mapped onto a $d+1$ classical NLSM
targeted on the projected space $CP^1=SU(2)/U(1).$ This mapping is fully controlled by $1/s$ expansion.
In $1d$, a  topological term also emerges to
discriminate between the spectrum of the low-energy excitations for integer and half-integer spin values $s$ \cite{haldane}.
This term produces an interference between the different topological sectors for
a half-integer $s$. As a result, gapless low energy excitations emerge out of that. For integer $s$, the topological term does not affect
the path integral and the model displays gapful excitations instead. Although the semiclassical approach is strictly speaking valid
for a large $s$ only, the qualitative distinction between the integer and half-integer chains holds true down to
the smallest possible spin magnitude, $s=1/2$. In other words the large-$s$ expansion gives a qualitatively correct picture  down to
the physical value of $s=1/2$.

To include doping, a natural guess might be that
a doped quantum AF model, in the low-energy quasiclassical limit,
would admit a mapping onto the $CP^{1|1}$ NLSM which in its turn can be viewed as a
natural extension of the $CP^1$ NLSM to incorporate extra fermionic degrees of freedom to
describe the doped holes.
The aim of the present paper is to show that this is not what happens for a doped quantum AF described by the $t-J$ model.
Because of the strong correlations inherent of the hopping lattice electrons due to a large on-site Coulomb repulsion,
strongly competing phases emerge at nonzero doping.
These are not captured by the semiclassics that substantially destroys strong correlations.
This in turn results in qualitatively different physics for large versus small values of the $su(2|1)$ representation index.

To set the stage, let us start with some definitions.
The $t-J$ model of correlated electrons is a lattice model
on the restricted $3^{\cal N}$ - dimensional electronic Hilbert space
(${\cal N}$-is a number of the lattice sites), where the occurrence of two electrons
on the same lattice site is strictly forbidden. This restriction comes from the strong
on-site Coulomb repulsion between the electrons which drive the strong correlations.
Explicitly, the $t-J$ model Hamiltonian
reads
\begin{equation}
H_{t-J}=-t\sum_{ij\sigma} \tilde{c}_{i\sigma}^{\dagger}
\tilde{c}_{j\sigma}+ J\sum_{ij} (\vec Q_i \cdot \vec Q_j -
\frac{1}{4}\tilde{n}_i\tilde{n}_j),
\label{01}\end{equation}
where $\tilde{c}_{i\sigma}=c_{i\sigma}(1-n_{i,-\sigma})$ is the Gutzwiller
projected electron operator (to avoid the on-site double
occupancy), $\vec
Q_i=\sum_{\sigma,\sigma'}\tilde{c}_{i\sigma}^{\dagger}\vec\tau_{\sigma\sigma'}\tilde{c}_{i\sigma'},
\,\vec\tau^2=3/4, $ is the electron spin operator and $\tilde
n_i=n_{i\uparrow}+n_{i\downarrow}-2n_{i\uparrow}n_{i\downarrow}$.
It contains the hopping term $\sim t$ and the spin exchange term $\sim J.$
The summation is extended
over the nearest neighbour (nn) sites of a $d$-dimensional bipartite lattice, $L=A\oplus B.$ Here if $i\in A$ then all its nn sites belong to the sublattice $B$.
For the hole-doped cuprates, $t\approx 0.5$ ev and $J\approx t/3.$

The implementation of the no double occupancy (NDO) constraint drives the system
towards the strong coupling regime\cite{epl}. This is a key point that makes the problem essentially non-perturbative.
Because of the NDO constraint, the hopping term cannot be taken as
a bare free interaction to derive a perturbation expansion in $J/t$. In fact, it can be diagonalized only in $1d$.
In higher dimensions, the hopping constrained electrons exhibit a nontrivial physics (e.g.,
the Nagaoka phase to describe a single hole doped into a $2d$ lattice of the hopping constrained electrons
\cite{nagaoka}.)
In the $t-J$ model, the effective charge and spin degrees of freedom are entangled  due to those strong electron correlations.

In view of the non-perturbative nature of strong electron correlations
any reliable  approach  is significantly important.
In particular, a considerable simplification  is provided by the
supersymmetric variant of the $t-J$ model. Namely,
in $d=1$, the $t-J$ model is exactly solvable at the supersymmetric (SUSY) point, $J=2t$ \cite{integr}.
At this point, it can be brought into a bilinear form
of the generators of the $su(2|1)$ superalgebra (also called $spl(2,1)$ in the literature) in the $3d$ fundamental degenerate irreducible representation (irrep).
Such a system exhibits a global $SU(2|1)$ supersymmetry which makes it solvable by Bethe ansatz in $1d$.
Away from the SUSY point, the symmetry group reduces to
$SU(2)\times U(1)$ which leads to the separate  conservation of the total
electron spin and the total number of doped vacancies as is physically appropriate for a doped quantum Heisenberg
model. Note that the $SU(2)\times U(1)$ group appears as an even subgroup of $SU(2|1)$.

A complementary theoretical method which is not based on a standard perturbation theory can also be established if we
explore the effects
of quantum and thermal fluctuations in the semiclassical limit.
It proves useful in describing the physics at large values of the relevant representation indices of a
global symmetry group.
A natural question then arises as to whether the $t-J$ model admits a reliable semiclassical treatment
that truly preserves strong correlations.
After all, this is precisely the case at half-filling $(\tilde n_i=1)$ at which the $t-J$ model reduces to the $SU(2)$ invariant
antiferromagnetic (AF) Heisenberg spin model.

The $su(2|1)$ algebraic approach provides two possible generalizations of the standard $t-J$ Hamiltonian to
include particles with spin higher than $1/2$, which is a necessary step to properly formulate a semiclassical expansion.
One possibility might be to consider $N$ electron orbitals at each site, which would correspond to the fundamental representation
of the $su(N|1)$ superalgebra instead of $su(2|1)$. One can then consider a fully antisymmetric (slave-boson) \cite{kotliar},
fully symmetric (slave-fermion), or  mixed ($L$-shaped) representations of $su(N|1)$ \cite{coleman}.
There appears an overall multiplier $\sim N$ in a corresponding path-integral action, which formally justifies a subsequent $1/N$
expansion. It is important to emphasize, however, that the uniform saddle point, whenever it is stable,
describes a Fermi liquid up to very small values of doping \cite{kotliar}. This is a qualitatively incorrect picture, since
at small doping strong correlation are known to display a manifestly non-Fermi liquid behaviour. Therefore the large-$N$ expansion
does not converge to the physical value $N=2.$
As a result, strong correlations are suppressed within the
proposed $1/N$ expansion.

An alternative procedure is to interpret the holes to be sites which have spin $s=q-1/2$ where $q\in {k}/2$ and $k$
is an integer.\cite{3}.
In view of that,
the sites without a "hole" acquire a spin $(q-1/2)+1/2=q.$ The latter possibility amounts to considering
the more general higher-spin $(q,q)$ representation of $su(2|1)$ rather than the $(q=1/2,q=1/2)$ fundamental one.
A  total action is proportional
to the representation index $q>>1,$ which  implies that a large-$q$ expansion may be applicable in that case.
However, there is a severe technical problem in explicitly carrying out the $1/q$ expansion.
In the $su(2|1)$ coherent-state path integral approach, the spin-charge entanglement is encoded in the
$SU(2|1)$ invariant path-integral measure: it does not admit a decomposition into
a product of a pure $SU(2)$ spin and $U(1)$ fermionic pieces.
The bosonic and fermionic fields are intertwined in the $SU(2|1)$ invariant measure in a
nontrivial manner which makes, except in a few trivial instances, an explicit computation of the pertinent path integral
rather problematic \cite{npb}.

To simplify the matter, we restrict ourselves to the case of the maximal possible global symmetry
of the $t-J$ Hamiltonian exhibited at the SUSY point, $J=2t$.
Although this simplifies the model considerably, strong electron correlations are still at work.
The spin-charge entanglement due to strong correlations manifests itself by means of
the even (representing spin) and odd (representing charge) generators
closed into a unique superalgebra, $su(2|1)$, so that
the charge and spin degrees of freedom transform themselves under the $SU(2|1)$ action through each other.
Technically, the semiclassical approach to the SUSY $t-J$ model follows the steps similar to those
encoded in the Haldane conjecture in dealing with the quasiclassical
Heisenberg $su(2)$ quantum spin model.
However, the $su(2|1)$ superalgebra brings into consideration a few qualitative
new features. In contrast with the $su(2)$
spin case for which any irrep is equivalent to its conjugate, the $su(2|1)$ superalgebra
admits two sets of inequivalent conjugate representations.
One can either place the $su(2|1)$ generators
in ${\it different} $ representations associated with two different sublattices, or use a single irrep all over
the full lattice.

We start with the semiclassical
effective theory for the SUSY spin model based on the so-called alternating group representations. In this case
the fundamental and the conjugate $su(2|1)$ representations are placed on different neighbouring sites.
Such a model has earlier been argued to produce  a canonical
$CP^{1|1}$ NLSM. It was written down in Ref.\cite{read} although
no explicit derivation based on a microscopical model was provided in their work.

Within the second option, we fix a classical ground state to be
realized by the generalized AF Neel superspin configuration. We explicitly derive the relevant
quasiclassical effective action to specify the entering coupling constants. Such a model
turns out to be given by the same  $CP^{1|1}$ NLSM. 

However both approaches are  shown to fail to accommodate the strongly correlated electron states  at finite doping.
As a result,
the physics exposed by the canonical NLSM
is unable to resolve
the theoretical challenge of properly describing strongly correlated metallic state in the vicinity of an insulating regime as a function of doping.
Although the present derivation may indeed be relevant for some physical models,
the generalized AF ansatz does not admit an interpretation in terms of the Hubbard operators at finite doping.
This by far rules out the possibility that
the canonical $CP^{1|1}$ NLSM may be considered as a relevant low-energy quasiclassical action to describe
the physics of the $t-J$ model at finite doping.

A novel point that helps us to derive the continuum limit of the SUSY $t-J$ model is that we express the lattice action entirely in terms of the projectors onto the underlying classical phase space -- the $su(2|1)$ coherent-state manifold. 
This allows us to simplify the calculations and in addition to provide a universal 
approach that produces a straightforward generalization to treat
other models with Hamiltonians constructed out of the generators of Lie (super)algebras. 
The overlap of the projectors is directly related to the important geometric structure -- the Kaehler (super)potential of the target (super)manifold. In the continuum limit, this overlap reduces to the metric function 
in terms of which the corresponding NLSM emerges.   
Such an approach is quite a general one:
It basically involves only one essential ingredient that determines the local geometry of the coherent-state manifold -- the  Kaehler (super)potential.

\section{Supersetting}

In this Section, we briefly review the basic facts concerning the $SU(2|1)$ supergroup
and the supersymmetric $t-J$ model.

\subsection{$su(2|1)$ superalgebra $\&$ $su(2|1)$ coherent states}

Our conventions for the $SU(2|1)$ supergroup are summarized as follows.
The $SU(2|1)$ supergroup in the fundamental representation
is the group of $(2+1)\times(2+1)$ unitary,
unimodular supermatrices with the Hermitian conjugate operation.
It is generated by even and odd generators, $\{B,Q_3,Q_{+},Q_{-}\}$ and
$\{W_{+},W_{-},V_{+},V_{-}\}$, respectively, which
satisfy the following
commutation rules \cite{ritten}:
\begin{eqnarray}
[Q_3,Q_{\pm}]=\pm Q_{\pm},[Q_{+},Q_{-}]=2Q_3,[B,Q_{\pm}]=[B,Q_3]=0, \cr
[B,V_{\pm}]=\frac{1}{2}V_{\pm},[B,W_{\pm}]=-\frac{1}{2}W_{\pm},
[Q_3,V_{\pm}]=\pm\frac{1}{2}V_{\pm},[Q_3,W_{\pm}]=\pm\frac{1}{2}W_{\pm},\cr
[Q_{\pm},V_{\mp}]=V_{\pm},[Q_{\pm},W_{\mp}]=W_{\pm},
[Q_{\pm},V_{\pm}]=[Q_{\pm},W_{\pm}]=0,\cr
\{V_{\pm},V_{\pm}\}=\{V_{\pm},V_{\mp}\}=\{W_{\pm},W_{\pm}\}=
\{W_{\pm},W_{\mp}\}=0,
\{V_{\pm},W_{\pm}\}=\pm Q_{\pm},\{V_{\pm},W_{\mp}\}=-Q_3\pm B.
\nonumber\end{eqnarray}
The second Casimir operator takes the form
$$K_2=\vec Q^2-B^2+\frac{1}{2}
(V_{+}W_{-}- V_{-}W_{+} +W_{+}V_{-} -W_{-}V_{+}).$$

Let $|b,q,q_3\rangle$ stand for a vector of any $su(2|1)$ abstract representation,
where $b,q$ and $q_3$ denote the eigenvalues of the operators
$B$ and $Q_3$, respectively, whereas $q$ is the quantum number that labels the eigenvalue of the $\vec Q^2$ which is actually $q(q+1)$.
A typical $SU(2|1)$ coherent state reads
\begin{eqnarray}
|z,\xi,\theta\rangle={\cal N}\,\exp( -\theta W_{-}
-\xi V_{-}+z Q_{-})|b,q,q\rangle,
\nonumber\end{eqnarray}
where $(z,\xi,\theta)\in SU(2|1)/U(1)\times U(1)$.
We are however interested
in the so-called atypical (degenerate) $b=q$ representation
that happens to be relevant for the $t-J$ model. This is specified by
\begin{eqnarray}
W_{-}|q,q,q\rangle=0
\nonumber\end{eqnarray}
and is called the $(q,q)$ representation with dimension $4q+1$.
This representation is spanned by $2q+1$ vectors
$\{|q,q,q_3\rangle,\quad -q\leq q_3\leq q\}$ of the even (bosonic)
and by $2q$ vectors $\{|q+1/2,q-1/2,q_3\rangle,\quad -q+1/2\leq q_3\leq q-1/2\}$
that correspond to the odd (fermionic) sectors.
Both the second and third order Casimir operators are zero in this
representation.
The coherent state reduces in the $(q,q)$ representation to
\begin{eqnarray}
|z,\xi\rangle=(1+|z|^2+\bar\xi\xi)^{-q}e^{-\xi V_{-}+z Q_{-}}|q,q,q\rangle,
\label{eq:2.1}\end{eqnarray}
where $z$ and $\xi$ are even and odd Grassmann parameters, respectively, to be
viewed upon as local coordinates on the supermanifold $CP^{1|1}=SU(2|1)/U(1|1).$
Classical images of the $su(2|1)$ generators are found to be
$(A^{c}\equiv\langle z,\xi|A|z,\xi\rangle)$:
\begin{eqnarray}
Q^{c}_3&=&-q(1-|z|^2)w,\quad (Q^{+})^{c}=2qzw, \quad (Q^{-})^{c}=2q\bar zw,\nonumber\\
B^{c}&=&q(1+|z|^2+2\bar\xi\xi)w,\quad (V^{+})^{c}=-2q z\bar\xi w, \quad
(V^{-})^{c}=2q\bar\xi w,\nonumber\\
(W^{+})^{c}&=&-2q\xi w,\quad  (W^{-})^{c}=-2q\bar z\xi w,\quad
w=(1+|z|^2+\bar\xi\xi)^{-1}.
\label{eq:2.2}\end{eqnarray}

There is also the second atypical $4q+1$-dimensional representation $(-q,q)$
which happens to be conjugate to the
$(q,q)$, i.e.,  $(q,q)^*= (-q,q).$  These irreps are inequivalent.

The $su(2|1)$ generators in the lowest atypical representation
 $(q=1/2,q=1/2)$ can be identified with the Hubbard operators \cite{wieg},
$X^{\sigma 0}_i$, defined by
$$X^{\sigma 0}_i=c^{\dag}_{i\sigma}(1-n_{i,-\sigma}),
\quad n_{i\sigma}n_{i,-\sigma}=0,$$
where $c_{i\sigma}$ is the annihilation operator of an electron at site $i$
with spin $\sigma=\pm$, and $n_{i\sigma}\equiv c^{\dag}_{i\sigma}c_{i\sigma}$.
In terms of these operators, the $t-J$ Hamiltonian becomes
\begin{eqnarray}
H_{t-J}&=&-t\sum_{ij\sigma}X^{\sigma 0}_iX^{0\sigma}_j +
J\sum_{ij}\left(\vec Q_i\vec Q_j-n_in_j/4\right)
\label{eq:3.1}\end{eqnarray}
with $J>0$ and the sum restricted to nearest neighbor sites.
Here the local electron spin operator
$$\vec Q_i=\frac{1}{2}\sum_{\sigma\sigma'}X^{\sigma0}_i
\vec{\tau}_{\sigma\sigma'}X^{0\sigma'}_i,$$ with
$\vec{\tau}=(\tau^1,\tau^2,\tau^3)$ being the Pauli matrices.
Electron on-site number operator $n_i=X^{++}_i+X^{--}_i.$

If $|0\rangle$ stands for a doped state (a hole) and $|\sigma\rangle$
denotes a state occupied by an electron with the spin projection $\sigma$,
Hubbard operators take the form
\begin{eqnarray}
X^{\sigma 0} =|\sigma\rangle\langle 0|,\quad X^{\sigma\sigma'}=
|\sigma\rangle\langle\sigma'|,
\label{eq:3.2}\end{eqnarray}
$X^{\sigma 0}$ being a fermionic operator and $X^{\sigma\sigma'}$
corresponding to the bosonic degrees of freedom.

It is clear that there are eight linearly independent operators since
by definition $$X^{00}+\sum_{\sigma}X^{\sigma\sigma}=I.$$
The $X$-operators
are closed into the $su(2|1)$ superalgebra
in the $3d$ degenerate $(1/2,1/2)$ representation in the following way,\cite{wieg}
$$Q_3=\frac{1}{2}(X^{++}-X^{--}),\quad Q_{+}=X^{+-},\quad Q_{-}=X^{-+},
\quad B=\frac{1}{2}(X^{++}+X^{--})+X^{00}$$
$$ V_{+}=X^{0-},\quad V_{-}=-X^{0+},\quad W_{+}=X^{+0},\quad
W_{-}=X^{-0}.$$
The even (bosonic) states $|1/2,1/2,1/2\rangle$ and $|1/2,1/2,-1/2\rangle$
are identified with the spin up and spin down states, $|+\rangle$ and
$|-\rangle$ , respectively, whereas the odd (fermionc) state $|1,0,0\rangle$,
with the doped state $|0\rangle$.
In contrast with the $(1/2,1/2)$ irrep,  the conjugate one, $(-1/2,1/2)$, does not provide a representation in terms of the
Hubbard operators.

\subsection{ SUSY $t-J$ model}

It turns out that at $J=2t$ this model exhibits in any dimension a
global $SU(2|1)$ supersymmetry.
This means that $H_{SUSY}=\sum_{ij}g_{\mu\nu}T^{\mu}_iT^{\nu}_j,
\,\,T^{\mu}\in su(2|1)$,
where $\nu,\mu=1,2,...,8,$ and summation over the group indices is implied.
It is crucial that $g_{\mu\nu}$ appears as the
$SU(2|1)$ invariant (under the adjoint action) tensor, with
$g_{\mu\nu}=str (T_{\mu}T_{\nu})$. Therefore any continuum version
of this interaction should also maintain that symmetry.

It can be shown that
\begin{equation}
H_{SUSY}= 2t\sum_{<ij>}\left[\vec Q_i\vec Q_j-B_iB_j+\frac{1}{2}
(V^{+}_iW^{-}_j- V^{-}_iW^{+}_j +W^{+}_iV^{-}_j -W^{-}_iV^{+}_j)\right]
\label{eq:4.1}\end{equation}
\begin{equation}
= t\sum_{<ij>}\left[X^{0-}_i X^{-0}_j +X^{0+}_i X^{+0}_j - X^{+0}_i X^{0+}_j
- X^{-0}_i X^{0-}_j +X^{+-}_i X^{-+}_j +X^{-+}_i X^{+-}_j- X^{++}_i X^{--}_j
- X^{--}_i X^{++}_j\right].
\label{eq:4.2}\end{equation}
Comparing eq.~(\ref{eq:4.1}) with the representation of the second Casimir
operator $K_2$ immediately proves that $H_{SUSY}$ displays global $SU(2|1)$
invariance.

This symmetry can be examined in terms of a more standard set of the $SU(2|1)$
group generators. Let us define
$$E^{\alpha\beta}=X^{\alpha\beta},\,\alpha\neq\beta,\quad E^{\alpha\alpha}=
(-1)^{|\alpha|}X^{\alpha\alpha}-1/2,\quad \alpha,\beta=\pm,0,$$
the $Z_2$ grading being defined so that
$|\alpha|=0$ for  $\alpha=\pm$ and $|\alpha|=1$ if $\alpha=0.$
With these definitions
\begin{equation}
H_{SUSY}=t\sum_{<ij>}E^{\alpha\beta}_iE^{\beta\alpha}_j(-1)^{|\beta|},
\label{eq:4.3}\end{equation}
which again coincides with the quadratic Laplace--Casimir element.

The coherent state in the (1/2,1/2) irrep takes the form
\begin{eqnarray}
|z,\xi\rangle=(1+|z|^2+\bar\xi\xi)^{-1/2}
(|+\rangle +\xi|0\rangle+z|-\rangle)
\label{eq:4.4}\end{eqnarray}
and the operator
$\hat Q(\bar z,\bar\xi;z,\xi)= |z,\xi\rangle\langle z,\xi|-1$,
in the $3d$ basis, $|\pm\rangle,\,|0\rangle$, displays the following matrix representation:
$$Q(\bar z,\bar\xi;z,\xi)=(1+|z|^2+{\bar\xi}\xi)^{-1}
\left(\begin{array}{ccc}-(|z|^2+\bar\xi\xi) &\bar z&-\bar\xi\\
z& -(1+\bar\xi\xi)&-z\bar\xi\\ -\xi&-\xi\bar z&-(1+2\bar\xi\xi+|z|^2)
\end{array}\right)$$
$$=\left(\begin{array}{ccc}Q_z-B &Q^-&-V^-\\
Q^+& -Q_z-B&V^+\\ W^+&W^-&-2B
\end{array}\right)^{c}$$
Note that $str\, Q=0$.
As a result,
\begin{equation}
\langle H_{SUSY}\rangle:=H^{c}_{SUSY}=t
\sum_{<ij>}Q^{\beta\alpha}_iQ^{\alpha\beta}_j(-1)^{|\beta|}=
t\sum_{<ij>}str (Q_iQ_j).
\label{eq:4.5}\end{equation}
Explicitly, we have that
\begin{equation}
H^{c}_{SUSY}=
t\sum_{<ij>}w_i^{-1}w_j^{-1}
\left[\bar z_iz_j+\bar z_jz_i-|z_i|^2-|z_j|^2-\xi_i\bar\xi_j(1+\bar
z_iz_j)-\xi_j\bar\xi_i(1+\bar z_jz_i)\right]=
t\sum_{<ij>}|\langle z_i\xi_i|z_j\xi_j\rangle|^2.
\label{eq:4.6}\end{equation}
where $w_i:=(1+|z_i|^2+\bar\xi_i\xi_i).$

In fact $|\langle z_i\xi_i|z_j\xi_j\rangle|^2 =str P_iP_j$, where the
projection operator $P_i=|z_i\xi_i\rangle\langle z_i\xi_j|$.
It can be checked that $str P=1$ and
\begin{equation}
\sum_{<ij>}str(Q_iQ_j)= \sum_{<ij>}str(P_iP_j) + const.
\label{eq:4.7}\end{equation}
The $str(P_iP_j)$ is directly related to
the Kaehler superpotential of the target superspace,
$$str(P_iP_j)=\exp{\Gamma_{ij}}, \quad\Gamma_{ij}= F_{ij}+F_{ji}-F_{ii}-F_{jj},$$
where the $CP^{1|1}$ Kaehler superpotential
$F_{ij}=\log(1+\bar z_iz_j+\bar\xi_i\xi_j).$  In the continuum limit
$F_{ij}+F_{ji}-F_{ii}-F_{jj}=\delta^2F_{ii}+o(a^2)$, $a$ being the lattice spacing,  and
with the second derivatives of $F$ emerging. Naturally from this definition those derivatives
manifest themselves as the entries of
the $CP^{1|1}$ metric in terms of which the nonlinear $CP^{1|1}$ sigma model is
going to be derived.

\section{$CP^{1|1}$ nonlinear sigma model: preliminary remarks}

Depending on a possible
choice of the classical ground state, one can arrive at different variants of
relevant  field theories to incorporate small quantum corrections.
To address a quasiclassical ($q>>1$) behavior of the SUSY $t-J$ model (\ref{eq:4.1})
one may adopt two possible ansatzes. First, since there are two inequivalent atypical representations
$(q,q)$ and $(-q,q)$, one may arrange the operators in (\ref{eq:4.1}) in an alternating manner:
the irrep $(q,q)$ of the $su(2|1)$ superalgebra
is placed on sites $i\in A$, whereas its conjugate, $(q,q)^*=(-q,q)$, are defined on sites $i\in B,\,\,$
with $L=A\oplus B.$ A tensor product of the two is decomposed
into a direct sum of the $SU(2|1)$ irreps that contains a singlet.
We briefly comment on this option at the end of the present Section.

Another route could be to assume that, classically (i.e., in the $q\to\infty$ limit),
the system minimizes its energy by making the nearest neighbour site
superspins acquire an "antiparallel"  orientation.
This option seems to formally provide
a direct extension of the Haldane ansatz for the $SU(2)$ spins to the $SU(2|1)$ generators.

Since in both cases our destination is the $CP^{1|1}$ sigma model, let  us keep in mind its
abstract definition.
A canonical nonlinear $CP^{1|1}$-valued $\sigma$-model can formally be described by a virtue of
the map $\varphi:\Sigma\to CP^{1|1}=SU(2|1)/U(1|1)$, where $\Sigma$ stands for
a space-time manifold with local coordinates $x^{\mu}, \mu=0,1,...,d\,
(x^0:=t),$
while the $CP^{1|1}$ denotes a $N=1$ superextension of a complex projecive
line $CP^{1}$ to the superspace with local coordinates $(z,\xi)=:\Phi^{\alpha},
\alpha=1,2.$ The K\"ahler potential $F=2q\log(1+\bar zz+{\bar\xi}\xi)$, with
$2q$ being integer, defines on $CP^{1|1}$ the Berry connection $a$ and
the Berry curvature $da$, where the external derivative
$d=\delta+\bar\delta$, $\delta=dz\partial_z+d\xi\partial_{\xi}$, and
$\bar\delta=d\bar z\partial_{\bar z}+d\bar\xi\partial_{\bar\xi}.$
(All derivatives are understood to be the left ones). Explicitly,
$a=\frac{i}{2}(\delta-\bar\delta)F=
\frac{i}{2}(dzF_z+d\xi F_{\xi}-d\bar zF_{\bar z}-d\bar\xi F_{\bar\xi})=
\frac{iq}{1+|z|^2+\bar\xi\xi}(\bar zdz-zd\bar
z-d{\bar\xi}\xi+\bar\xi d\xi) =:a_zdz+a_{\bar z}d\bar
z+d\xi a_{\xi}+a_{\bar\xi}d\bar\xi$. The irrep $(1/2,1/2)$ is specified by
setting $q=1/2$. The pull-back of the
one-form $a$ is given by $\varphi^* a,\quad (\varphi^*a)_{\mu}=
a_z\partial_{\mu}z+...$

Under the $SU(2|1)$ action, $F\to F+\Lambda(z,\xi)+
\overline{\Lambda(z,\xi)}$, which means that the symplectic one-form $a$
undergoes the $U(1)$ gauge transformation,
$$a\to a+d\psi, \quad
\psi=\frac{i}{2}(\Lambda-\bar\Lambda).$$ The associated invariant canonical supersymplectic
two-form is
\begin{equation}
\Omega^{(2)}=d{a}=-i\delta\bar\delta F=-i(F_{z\bar
z}dz\wedge d\bar z
-F_{z\bar\xi}dz\wedge d\bar\xi-F_{\xi\bar z}d\xi\wedge
d\bar z+F_{\xi\bar\xi}d\xi\wedge d\bar\xi),
\label{eq:5.0}\end{equation}
where $$ F_{z\bar
z}=\frac{2q}{(1+|z|^2+{\bar\xi}\xi)^2}(1+\bar\xi\xi),\quad
F_{z\bar\xi}=-\bar z\xi\,\frac{2q}{(1+|z|^2+{\bar\xi}\xi)^2},\quad
F_{\bar z\xi}=\bar\xi z\,\frac{2q}{(1+|z|^2+{\bar\xi}\xi)^2},$$
$$F_{\xi\bar\xi}=-\frac{2q}{1+|z|^2},$$
belong to the exterior algebra on $CP^{1|1}$. The latter
is a bi-graded $Z\times Z_2$ algebra, where the $Z$-gradation is the usual
gradation of de Rham complexes, while $Z_2$-gradation is a natural
gradation of Grassmann algebra. For any two superforms $\beta_1$ and
$\beta_2$ on $CP^{1|1}$, one has
$\beta_1\wedge\beta_2=(-)^{a_1a_2+b_1b_2}\beta_2\wedge\beta_1$, where
$a_i(b_i)$ is the degree of $\beta_i$ with respect to the $Z(Z_2)$
gradation. Hence, $dz\wedge d\bar z=-d\bar
z\wedge dz,\,dz\wedge d\bar\xi=-d\bar\xi\wedge dz,\,d\xi\wedge
d\bar\xi=d\bar\xi\wedge d\xi.$ In terms of the coherent states
$$a=i\langle z,\xi|d|z,\xi\rangle,\quad da= id\langle z,\xi|d|z,\xi\rangle =
i\, str(P\, dP\wedge dP).$$

In view of this the supersymplectic two-form  can be written in the form
$$\Omega^{(2)}=-ig_{\alpha\bar\beta}d\Phi^{\alpha}\wedge
d\bar\Phi^{\beta},\quad \Phi =(\Phi^1,\Phi^2):=(z,\xi),$$
and the $SU(2|1)$ invariant metric supertensor then reads,
\begin{equation}
g(\bar\Phi,\Phi)= g_{\alpha\bar\beta}d\Phi^{\alpha}d\bar\Phi^{\beta}
=F_{z\bar z}dzd\bar z-F_{z\bar\xi}dzd\bar\xi-F_{\xi\bar z}d\xi
d\bar z+F_{\xi\bar\xi}d\xi d\bar\xi,
\label{eq:metric}\end{equation}
where differentials
of the odd coordinates are understood to anticommute among themselves,
and commute with differentials of the even coordinates, with the latter
exhibiting the usual behavior.

The action functional for the $SU(2|1)$ invariant $\sigma$-model can then be written
in the form $(D=1+d)$:
\begin{eqnarray}
&&S_g=-\frac{1}{2g^2}\int_{\Sigma}d^{D}x\,str(\varphi^*g)=
\nonumber \\
&&-\frac{1}{2g^2}\int_{\Sigma}d^{D}x\left[\partial_{\mu}\Phi\,g
\partial_{\mu}\bar\Phi\right]=
-\frac{1}{2g^2}\int_{\Sigma}d^{D}x\left
[F_{z\bar z}\partial_{\mu}z\partial_{\mu}\bar z-F_{z\bar\xi}\partial_
{\mu}z\partial_{\mu}\bar\xi  -F_{\xi\bar z}\partial_{\mu}\bar
z\partial_{\mu}\xi
+F_{\xi\bar\xi}\partial_{\mu}\xi\partial_{\mu}
\bar\xi \right]=\nonumber \\
&&-\frac{1}{2g^2}\int_{\Sigma}d^{D}x\left[\frac{1}{(1+|z|^2+{\bar\xi}\xi)^2}
\left((1+\bar\xi\xi)\partial_{\mu}z\partial_{\mu}\bar z
+\bar z\xi\partial_{\mu}z\partial_{\mu}\bar\xi-
z{\bar\xi}\partial_{\mu}\bar
z\partial_{\mu}\xi\right)-\frac{1}{1+|z|^2}\partial_{\mu}\xi\partial_{\mu}
\bar\xi\right],
\label{eq:5.1} \end{eqnarray}
where the space-time $\Sigma$ is assumed to be a flat euclidean manifold.

It is also well known that a topological (metric independent) $SU(2|1)$ invariant
Wess--Zumino term can be added to $S$.
In general, it can be defined as follows. Consider a map $\varphi:\Sigma\to M$,
of a $D$ space-time surface $\Sigma$ into
an ordinary manifold (not a supermanifold) $M$. Let us
assume that a closed $D+1$-form, $\Omega^{(D+1)},$ with $d
\Omega^{(D+1)} =0,$ can be picked up on $M$. In case it defines a
trivial cohomology class, one has globally $\Omega^{(D+1)}
=d\Omega^{(D)}$ for a certain $D$-form on $M$. In that case one
simply defines $$S_{WZ}=\int_{\Sigma}\varphi^*\Omega^{(D)}.$$ If, however,
$\Omega^{(D+1)}$ is not globally exact, one may either consider
a set of coordinate patches $\Sigma_a$, so that $\Sigma=\cup_a\Sigma_a$,
and a field $\Omega^{(D)}_a$ defined in each patch such that
$\Omega^{(D+1)}=d\Omega^{(D)}_a$, or one may try to extend $\varphi$ to
a certain $\tilde\varphi:\, \tilde\Sigma\to M$, such that
$\partial{\tilde\Sigma}=\Sigma$ and, consequently
$$S_{WZ}=\int_{\tilde\Sigma}\tilde\varphi^*\Omega^{(D+1)}.$$
In both cases the theory is well-defined, provided
$\Omega^{(D+1)}$ defines an integral cohomology class, i.e., with the period of
the form being an integer number,
$$ \frac{1}{2\pi}\int_{V_{D+1}}\Omega^{(D+1)}=N,$$ where $V_{D+1}$ stands
for a $D+1$ oriented closed manifold in $M$.

The generalization of the above formalism to the supersetting is however not
straightforward.
First, any superform that involves
$d\xi$ or/and $d\bar\xi$ defines a trivial cohomology. This observation
seems to simplify the consideration.  What makes it more
involved is the fact that superforms cannot in general be integrated. This might
be a sort of disaster, if it were not for the fact that there
is a natural mapping $\varphi:\Sigma\to CP^{1|1}$, As a matter of fact,
relevant integrals can be defined via the pull-back $\varphi^*$:
\begin{equation}
\int_{\varphi(\Sigma)}d\xi\wedge d\bar\xi:=
\int_{\Sigma}\varphi^*(d\xi\wedge d\bar\xi)=
\int_{\Sigma}\partial_{\mu}\xi\partial_{\nu}\bar\xi\, dx^{\mu}\wedge
dx^{\nu}.
\label{definition}\end{equation}

For example, the $2d$ $t-J$ continuum model can be viewed upon as a mapping
$\varphi:(x,t)=S^2\to CP^{1|1}$, with the space-time $(x,t)$ being compactified
into $S^2$. We are therefore looking for a closed $SU(2|1)$-invariant
$3$-forms on $CP^{1|1}$ so that $\Omega^{(3)}=d\Omega^{(2)}$ globally.
One easily finds that there exists four trivially closed forms on
$CP^{1|1}$ generated by the functions $F_{z\bar z\xi}+F_{z\bar
z\bar\xi},\,\,F_{\xi\bar\xi z}+F_{\xi\bar\xi\bar z} \,\, F_{z\bar
z\xi}+F_{\bar z\xi\bar\xi}$ and $F_{z\bar\xi\xi}+F_{z\bar z\bar\xi}$,
respectively. These in turn yield
$$\Omega^{(2)}_1=-iF_{z\bar z}dz\wedge d\bar z,\quad
\Omega^{(2)}_2=-iF_{\xi\bar\xi}d\xi\wedge d\bar\xi,\quad
\Omega^{(2)}_3=-iF_{z\bar\xi}dz\wedge d\bar\xi,\quad
\Omega^{(2)}_4=-iF_{\bar z\xi}d\bar z\wedge d\xi.$$
The only $SU(2|1)$-invariant combination of all these choices is the symplectic
superform $\Omega^{(2)}$.

As a result, the Wess-Zumino action takes the form
\begin{equation}
S_{WZ}=: S_{\theta}=(i\theta/4\pi q)\int_{CP^{1|1}}\Omega^{(2)},
\label{eq:theta}\end{equation}
with the coefficient $\theta$ of the topological term
being determined at microscopical level.  Explicitly, this reads
\begin{eqnarray}
S_{\theta}&=& (\theta/2\pi)\int_{\Sigma}\frac{dx^{\mu}\wedge
dx^{\nu}} {(1+|z|^2+{\bar\xi}\xi)^2}\left[
(1+{\bar\xi}\xi) z_{\mu}\bar
z_{\nu}+ z\bar\xi\bar z_{\mu}\xi_{\nu}-\bar z\xi z_{\mu}\bar\xi_{\nu}\right]
\nonumber\\
&&+(\theta/2\pi)\int_{\Sigma}\frac{dx^{\mu}\wedge dx^{\nu}}
{1+|z|^2}\xi_{\mu}\bar\xi_{\nu}.
\label{eq:5.2}\end{eqnarray}

If one discards the fermionic degrees of freedom, thereby considering a
conventional $2d$ $\sigma$-model as a mapping $\varphi:CP^1\to CP^1$,
the above Wess-Zumino term reduces to a topological invariant \cite{3d}
$$ S_{\theta}\to S_{\theta}
=(\theta/2\pi)\int_{CP^1}\frac{dx^{\mu}\wedge dx^{\nu}}
{(1+|z|^2)^2}z_{\mu}\bar z_{\nu}=(\theta/2\pi)\int_{CP^1}
\varphi^*\frac{dz\wedge d\bar z}{(1+|z|^2)^2}$$
$$= (\theta/2\pi)(deg\phi)\int_{CP^1}\frac{dz\wedge d\bar z}{(1+|z|^2)^2}
=i\theta N,$$
using
$$i\int_{CP^1}\frac{dz\wedge d\bar z}{(1+|z|^2)^2} =2\pi.$$
At $\theta=2\pi q$ this is the familiar Haldane's result for the
topological term of the spin-q $2d$ continuum AF action \cite{1}.

As a result of that, the total $SU(2|1)$ invariant action takes the form
\begin{eqnarray}
&&S=S_g +S_{\theta}=
-\frac{1}{2g^2}\int_{\Sigma}d^2x\,str(\varphi^*g)+
(i\theta/4\pi q)\int_{CP^{1|1}}\Omega^{(2)}=
\nonumber \\
&&-\frac{1}{2g^2}\int_{\Sigma}d^2x\left[\frac{1}{(1+|z|^2+{\bar\xi}\xi)^2}
\left((1+\bar\xi\xi)\partial_{\mu}z\partial_{\mu}\bar z
+\bar z\xi\partial_{\mu}z\partial_{\mu}\bar\xi-
z{\bar\xi}\partial_{\mu}\bar
z\partial_{\mu}\xi\right)-\frac{1}{1+|z|^2}\partial_{\mu}\xi\partial_{\mu}
\bar\xi\right]\nonumber\\
&&+ (\theta/2\pi)\int_{\Sigma}\frac{dx^{\mu}\wedge
dx^{\nu}} {(1+|z|^2+{\bar\xi}\xi)^2}\left[
(1+{\bar\xi}\xi) z_{\mu}\bar
z_{\nu}+ z\bar\xi\bar z_{\mu}\xi_{\nu}-\bar z\xi z_{\mu}\bar\xi_{\nu}\right]
+(\theta/2\pi)\int_{\Sigma}\frac{dx^{\mu}\wedge dx^{\nu}}
{1+|z|^2}\xi_{\mu}\bar\xi_{\nu}.
\label{eq:5.3} \end{eqnarray}

The $2D$ NLSM given by Eq.(\ref{eq:5.3}) 
has been argued to yield a continuum limit of the alternating $su(2|1)$ quantum superspin chain
with alternating fundamental and conjugate fundamental representations \cite{read}.
In this reference, the action (\ref{eq:5.3}) is written down through the homogeneous (gauge dependent) coordinates,
whereas we employ the inhomogeneous variables. They are related to each other by a change of variables to explicitly
resolve the gauge redundancy.

The $SU(2|1)$ alternating chain is closely related
with the properties of the percolation cluster boundaries. It is gapless, and flows to the fixed point
that is described by a logarithmic conformal field theory at central charge $c=0.$ This theory corresponds
to the strong coupling regime of a $CP^{1|1}$ sigma model \cite{read}.
However, the model (\ref{eq:5.3}) cannot describe strongly correlated electrons. The point is that
strong electron correlations are encoded into the properties of the
fundamental $(1/2,1/2)$ irrep of $su(2|1)$. Generators of this representation can be identified
with the Hubbard operators (\ref{eq:3.1}). In contrast,
the conjugate irrep $(-1/2,1/2)$ of the $su(2|1)$ superalgebra does not admit
a description in terms of the Hubbard operators.
In particular, the slave-fermion representation of the $(-1/2,1/2)$  generators necessarily involves
"fermions" that anticommute to $-1$: $\{f_i,f_i^+\}=-1,\,i\in B.$ Physically, those fermions cannot
represent doped holes.
Consequently, the $(-1/2,1/2)$ representation cannot accommodate
physical electrons at any finite doping.

\section{$CP^{1|1}$ nonlinear sigma model: explicit derivation}

Let us now turn to our second option. In dealing with a quasiclassical limit of the
SUSY $t-J$ model (\ref{eq:4.1}) we now start with a generalized AF ansatz. It implies that on both even and odd sublattices
the superspins predominantly point in opposite "directions". In the quasiclassical limit, $q\to\infty$,
this is supposed to yield  an exact classical ground state, whereas in its vicinity one can derive a low-energy effective action
to incorporates small quantum/thermal fluctuations.
Such a theory is supposed to be fully controlled by a $1/q$ expansion.

As far as we know, a relevant theory has never been derived explicitly in this case. Presumably it is due to the fact
that the standard Haldane's
ansatz $\vec S^c_i\to -\vec S^c_i, \, i\in B$ (see Appendix A) cannot be generalized straightforwardly to the superspins.
There is no such change of variables in a relevant $SU(2|1)$ coherent-state path-integral action that can change the sign of
all the superspins simultaneously on one of the sublattices.

To proceed with that, we employ instead the projectors on the coherent-state manifold. 
This enables us to rewrite the lattice action entirely in terms of the projectors:  $H_{SUSY}^{cl}=str\sum_{ij}P_iP_j$ and $a_i =i\,str(dP_i)$. 
This simplifies our calculations greatly providing in addition a universal approach
to quasiclassically treat the model Hamiltonians which display global invariance under  a Lie (super)group action. 
The overlap of the projectors is directly related to the important geometric structure -- the Kaehler (super)potential of the target (super)manifold. In the continuum limit, this overlap reduces to the metric function in terms of which emerges the NLSM in question.

In case there is projector $P_i$
on the sublattice $A$, on sublattice B one has to change
$P_i\to 1-P_i, \, i\in B$. Indeed, $Q^c=str QP \to str Q(1-P)=-Q^c.$
Assuming the semi-classical Neel superspin configuration, we can then make the ansatz
\begin{equation}
H^{c}_{SUSY}=\sum_{<ij>}str(P_iP_j) \to
\sum_{<ij>}str(1-P_iP_j)=\sum_{<ij>}(1-\exp{\Gamma_{ij}}).
\label{eq:4.8}\end{equation}
As we show in the Appendix A, this procedure at half-filling is equivalent to the Haldane ansatz for the AF Heisenberg model.
It then follows that the continuum limit to describe the theory around a
classical Neel state is determined by only one ingredient,
the $SU(2|1)$ covariant Kaehler superpotential.
Away from the SUSY point, this approach does not hold, however.

To explicitly derive the relevant sigma model let us start by writing down a lattice
Euclidean action that enters the path-integral representation of the $t-J$
partition function,
$$S=i\sum_i\oint a_i-\int H_{SUSY}^{cl}(t)dt.$$
The first piece in this action is a topological term: it emerges as a sum of the
line integrals of the site dependent $su(2|1)$ symplectic one-forms
$a_i=i\langle z_i,\xi_i|d|z_i,\xi_i\rangle$ and involves no metric.
It is commonly referred to as the Berry phase term or, the Wess-Zumino term
or, the 1D Chern-Simons term. The second part is a metric dependent term.

Let us first turn to the evaluation of a continuum form of
that second part of the action.
For a general magnitude of the representation index q we have
$H_{SUSY}^{cl}=2q\,H^{cl}_{q=1/2},$ with parameter $t$ being replaced
by $t\to t/2q$.
Low-energy continuum limit then implies that $2q\to\infty$.
In view of eqs.~(\ref{eq:4.5},\ref{eq:4.8}) we get in $1D$
\begin{equation}
H^{cl}_{q=1/2}=:H^{cl}=-t\sum_i\Gamma_{ii+1} +o(a^2),
\label{eq:6.1}\end{equation}
where $a$ stands for the lattice spacing.
Here $\Gamma_{ii+1}:=F(\bar z_i,\bar\xi_i|z_{i+1},\xi_{i+1})+
F(\bar z_{i+1},\bar\xi_{i+1}|z_i,\xi_i)
-F(\bar z_i,\bar\xi_i|z_i,\xi_i)
-F(\bar z_{i+1},\bar\xi_{i+1}|z_{i+1},\xi_{i+1})$ and
the Kahler potential $F(\bar z_i,\bar\xi_i|z_j,\xi_j)=\log(1+\bar z_iz_j+
\bar\xi_i\xi_j)$.

To proceed, we impose the ansatz
\begin{equation}
z_i\to z_i+\delta z_i,\quad \xi_i\to \xi_i+\delta\xi_i,\quad i\in A;\qquad
z_i\to z_i-\delta z_i,\quad \xi_i\to \xi_i-\delta\xi_i,\quad i\in B,
\label{eq:6.2}\end{equation}
where $\delta z_i\sim a,\,\delta\xi_i\sim a$ are small AF field fluctuations.
In the continuum limit, $a\to 0$, eq.~(\ref{eq:6.1}) takes the form
\begin{equation}
H^{cl}=t\sum_ia^2(\Phi_i'\,g(\bar\Phi_i,\Phi_i)\,\bar\Phi'_i)+
4t\sum_i(\delta\Phi_i\, g(\bar\Phi_i,\Phi_i)\,\delta{\bar\Phi}_i) +o(a^2),
\label{eq:6.3}\end{equation}
where the superfields $\Phi=(z,\xi),\,\bar\Phi^t=(\bar z,\bar\xi),$ and
$g$ is the $SU(2|1)$ invariant metric supertensor explicitly
given by~(\ref{eq:metric}).

Let us now turn to the topological term.
As already mentioned, on sublattice B we let $P\to 1-P$. Because of this
the one-form $a_i$ changes its sign on sublattice B.
Hence we get
\begin{equation}
i\sum_i\oint (a_i+\delta a_i)=
i\sum_i\oint a_i=i\sum_{i\in A}\oint a_i - i\sum_{i\in B}\oint a_i
+ i\sum_{i\in A}\oint \delta a_i - i\sum_{i\in B}\oint \delta a_i,
\label{eq:6.3'}\end{equation}
where variation $\delta a_i$ is produced by change~(\ref{eq:6.2}).
The second piece of eq.~(\ref{eq:6.3'}) that arises from
the changes of the Berry phase in each of the sublattices yield
$$i\sum_{i\in A}\oint \delta a_i - i\sum_{i\in B}\oint \delta a_i
  = \sum_i\int dt\left[\delta\Phi_i g(\bar\Phi_i,\Phi_i)\dot{\bar\Phi}_i
-\dot\Phi_i g(\bar\Phi_i,\Phi_i) \delta\bar \Phi_i\right].$$
This term along with eq.~(\ref{eq:6.3}) give the following contributions to the
action
$$ \sum_i\int dt\left[(\delta\Phi_i g(\bar\Phi_i,\Phi_i)\dot{\bar\Phi}_i)
-(\dot\Phi_i g(\bar\Phi_i,\Phi_i) \delta\bar \Phi_i)+
ta^2(\Phi_i'\,g(\bar\Phi_i,\Phi_i)\,\bar\Phi'_i)+
4t(\delta\Phi_i\, g(\bar\Phi_i,\Phi_i)\,\delta{\bar\Phi}_i)\right] +o(a^2).$$
Extremizing the action with respect to $\delta\Phi$ and $\delta{\bar\Phi}$
produces the metric dependent contribution
$$-q\sum_i\int dt\left(\frac{\dot\Phi_i\,g_i\dot{\bar\Phi}_i}{4qt}+a^24qt
\Phi'_i\,g_i\,\bar\Phi'_i\right).$$ Note that we have restored the full $q$
dependence.  In the continuum limit this contributes to the total action in the
form
\begin{equation}
S_g=-\frac{1}{2g^2}\int dxdt\left(\frac{\dot\Phi g(\bar\Phi,\Phi)
\dot{\bar\Phi}}{c}+c\Phi'g(\bar\Phi,\Phi)\bar\Phi'\right),\qquad g^2=1/q,\quad
c=4qat.
\label{eq:6.3''}\end{equation}
At $c=1$ eq.~(\ref{eq:6.3''}) goes over into the
Lorentz invariant action targeted on $CP^{1|1}$:
\begin{eqnarray}
S_g&=&-\frac{1}{2g^2}\int dxdt\left(\partial_{\mu}\Phi g(\bar\Phi,\Phi)
\partial_{\mu}{\bar\Phi}\right)\nonumber\\
&&-\frac{1}{2g^2}\int dxdt\left
[F_{z\bar z}\partial_{\mu}z\partial_{\mu}\bar z-F_{z\bar\xi}\partial_
{\mu}z\partial_{\mu}\bar\xi  -F_{\xi\bar z}\partial_{\mu}\bar
z\partial_{\mu}\xi
+F_{\xi\bar\xi}\partial_{\mu}\xi\partial_{\mu}
\bar\xi \right],
\label{eq:6.3'''}\end{eqnarray}
which exactly coincides with Eq.~(\ref{eq:5.1}) if we take $g^2=1/q.$

Let us now focus on the evaluation of the topological invariant $S_{\theta}$
that arises from the first piece of the variation of the Berry phase term,
$$\oint a_i-\oint a_{i+1}=:\delta\oint a_i.$$  Here a small
change in the integral is due to the following transformations of the
coordinates:
\begin{equation}
\delta z_i=z_{i+1}-z_i=z(x_i+a)-z(x_i)=az'(x_i)+O(a^2);\quad
\delta\xi_i=a\xi'(x_i)+O(a^2).
\label{eq:6.4}\end{equation}
This transformation is generated by the infinitesimal vector field
$$\delta X=\delta z\partial_z+ \delta\bar z\partial_{\bar z}+\delta\xi
\partial_{\xi}+\delta{\bar\xi}\partial_{\bar\xi}.$$
As is well known a variation of a line integral due to a local one-parameter
group of diffeomorphisms generated by a vector field $\delta X$
gives
\begin{equation}
\delta\oint a_i=\oint {\cal L}_{\delta X_i}a_i,
\label{eq:6.5}\end{equation}
where ${\cal L}_ X$ stands for a Lie derivative with respect to the vector
field $X$. By making use of the identity
$${\cal L}_X=i_X\,d+d\,i_X$$ one can then transform~(\ref{eq:6.5}) into
\begin{equation}
\delta\oint a_i=\oint i_{\delta X_i}\,da_i +\oint d\,(i_{\delta X_i}a_i),
\label{eq:6.6}\end{equation}
where $i_X$ denotes the interior product. The last term in~(\ref{eq:6.6}) drops
out thanks to Stokes' theorem and we are left only with the first term.
To evaluate it we need to know how the interior derivative acts on
$da$ given by eq.~(\ref{eq:5.0}).
Explicitly we get
$$i_{\delta X_i}(F_{\bar z_iz_i}dz_i\wedge d\bar z_i)=
F_{\bar z_iz_i}(\delta z_i d\bar z_i -dz_i\delta\bar z_i),\quad
i_{\delta X_i}(F_{z_i\bar\xi_i}dz_i\wedge d\bar\xi_i)= F_{z_i\bar\xi_i}
(\delta z_i d\bar\xi_i-dz_i\delta\bar\xi_i),$$
$$i_{\delta X_i}(F_{\xi_i\bar z_i}d\xi_i\wedge d\bar z_i)
= F_{\xi_i\bar z_i}(\delta\xi_i d\bar z_i-d\xi_i\delta\bar z_i),\quad
i_{\delta X_i}(F_{\xi_i\bar\xi_i}d\xi_i\wedge d\bar\xi_i)=
F_{\xi_i\bar\xi_i}(\delta\xi_i d\bar\xi_i-d\xi_i\delta\bar\xi_i).$$
Here $\delta z_i=z'_ia,\,dz_i=\dot z_i dt$ and
similarly for the variations of $\xi_i$.
The first equation tells us that in the continuum limit
$$\sum_iF_{\bar z_iz_i}(\delta z_i d\bar z_i -dz_i\delta\bar z_i)=
\int F_{\bar zz}(z'\dot{\bar z}-\dot z\bar z')dx\wedge dt=\int
F_{\bar zz}dz\wedge d\bar z ,$$
where $z=z(x,t)$ and $dz=z'dx+\dot zdt.$
Similar relations hold for the remaining parts of the expressions and we finally
end up with the following contribution to the action coming from the first
piece of the topological term
\begin{equation}
S_{\theta}=i\sum_{i\in A}\oint a_i - i\sum_{i\in B}\oint a_i
=\frac{i}{2}\int_{CP^{1|1}} \Omega^{(2)}.
\label{eq:6.7}\end{equation}
Comparing this with eq.~(\ref{eq:theta}), gives $\theta=2\pi q.$
Note that factor $1/2$ arises because of the doubling of the lattice spacing.

In summary, we have derived the $2D$ $CP^{1|1}$ NLSM
to describe
the low--energy physics of the $1d$ quantum $su(2|1)$ superspin
chain~(\ref{eq:4.2}) in the form
$$S=S_g+S_{\theta}$$
\begin{eqnarray}
=-\frac{1}{2g^2}\int dxdt[F_{z\bar z}\partial_{\mu}z\partial_{\mu}\bar z-F_{z\bar\xi}\partial_
{\mu}z\partial_{\mu}\bar\xi  -F_{\xi\bar z}\partial_{\mu}\bar
z\partial_{\mu}\xi
+F_{\xi\bar\xi}\partial_{\mu}\xi\partial_{\mu}
\bar\xi]
+
\frac{i\theta}{4\pi q}\int_{CP^{1|1}} \Omega^{(2)}, \quad
g^2=1/q,\quad\theta=2\pi q,
\label{eq:6.8}\end{eqnarray}
which agrees with eq.~(\ref{eq:5.3}),
the phenomenological parameters being now explicitly specified.
In particular, the $\theta$- term ensures that the model (\ref{eq:6.8}) contains low-energy gapless excitations in
accordance with the exact solution for the alternating superspin chain obtained by Bethe ansatz \cite{integr1}.

However, once again, the ansatz (\ref{eq:4.8}) cannot describe
the Hubbard operators on both sublattices simultaneously as given by Eq.(\ref{eq:4.2}).
On the sublattice $B$, we get $B_i^c\to -B_i^c$, which is inconsistent with the definition of the $su(2|1)$ generator $B$
in terms of Hubbard operators:
$B=\frac{1}{2}(1+X^{00})$. The eigenvalues of the operator $X^{00}$ are either $0$ and $1$ and they describe either an absence of the doping
or a doped hole, respectively. This
change of sign of $B$ implies that the doping becomes "negative" and this is physically inappropriate. At half-filling, however, the operator $B$ becomes a c-number 
that takes on the values of +q and -q on the two different sublattices. This does not 
affect the effective action that reduces in this case to the $CP^1$ NLSM.   
However, in the presence of a nonzero doping 
the proper quasiclassical ground state has a more complicated
structure than that displayed by a pure Neel state. As a result, the Neel state is not a good reference state, in those cases.

As a matter of fact, an exact diagonalization of a two-site  cluster
indicates that the AF ground state of the  SUSY $t-J$ model (\ref{eq:4.2}) is a quartet, rather than a Neel state or a singlet \cite{foerster}.
This result can be visualized in terms of the following $su(2|1)$ diagram in the Clebsch-Gordon series
$3\otimes 3=5\oplus 4.$ 
At half filling, this quartet reduces to a $SU(2)$ spin singlet. It is quite a
nontrivial problem to figure out whether a proper large-$q$ generalization of such a state may serve as a reference state for the $CP^1$ Haldane action, which is a necessary condition for the quartet state to represent a proper reference state away from half filling.

Another alternative possibility might be to consider the low energy effective action of the $t-J$ model 
as being represented by the  spinless fermions  interacting with the $CP^1$ purely spin sigma model via the emergent gauge field.
The gauge field appears as a U(1) connection of the corresponding magnetic monopole bundle \cite{npb}. At half-filling, such a theory automatically reduces to the $CP^1$  NLSM as required. However, the emergent $SU(2|1)$ invariant measure intertwines the spin and charge degrees of freedom in a highly nontrivial manner, which is again a manifestation of strong correlations. As a result, no nontrivial computations are available so far within this approach.  

However, even in the case one happens to guess correctly an appropriate reference state there is still an issue as to  whether  strong correlations that are built into the theory at $q=1/2$ survive at the large -$q$ limit.
Our guess is that it does not persist. 
Indeed, a state with one hole and a state with two holes acquire spins $q-1/2$ and $q-1$, respectively. Those states can hardly be distinguished from each other in the large-$q$ limit.
The crucial no doubly occupied constraint 
that encodes the essence of strong correlations seems to be relaxed in this limit.

\section{Conclusion}

We consider the $t-J$ model of strongly correlated electrons at the SUSY point, $J=2t$.
The constrained electron operators that describe the correlated electrons
can be identified with the generators of the fundamental $(1/2,1/2)$ $3d$ representation
of the $su(2|1)$ superalgebra given by the canonical Hubbard operators.
However, to investigate the issue as to whether the quasiclassics may capture or not
the physics of strong correlations in a reliable way,
we make use of the higher $(q,q)$ representation  for large magnitudes of the superspin $q$. Starting from the generalized AF Neel configuration
of the superspins on a bipartite lattice, we derive the low-energy effective action -- the canonical $CP^{1|1}$ NLSM.

We then show that although this derivation may indeed be relevant for some physical models,
the generalized AF ansatz does not admit an interpretation in terms of the Hubbard operators at finite doping.
This by far rules out the possibility that
the canonical $CP^{1|1}$ NLSM may be considered as a relevant low-energy quasiclassical action to describe
the physics of the $t-J$ model at finite doping.

The proposed derivation of the NLSM is based on the representation of a classical image of the SUSY
$t-J$ model Hamiltonian in terms of the projectors on the $SU(2|1)$ coherent-state manifold.
Such an approach appears as quite a general one:
It basically involves only one essential ingredient that determines a local geometry of the 
coherent-state  manifold -- a  Kaehler (super)potential.

\section{Appendix A}

Let us show how this technique works in the case of the AF Heisenberg model.

The $su(2)$ algebra is spanned by the operators $S^{+},S^{-},S_z$ with the commutation
relations $[S_z,S^{\pm}]=\pm S^{\pm}, [S^{+},S^{-}]=2S_z$. In the fundamental
representati $(s=1/2),$ we define
$$S^+=|+\rangle\langle -|,\,
S^-=|-\rangle\langle +|,\,
S_z=\frac{1}{2}(|+\rangle\langle
+|-|-\rangle\langle -|),$$
with the $|\pm\rangle$ being the spin-up and spin-down states, respectively.

The $su(2)$ normalized coherent state in this irrep is \cite{stone89}
$$|z\rangle =(1+|z|^2)^{-1/2}e^{zS^{-}}|+\rangle=(1+|z|^2)^{-1/2}
(|+\rangle+z|-\rangle),$$ where $z$ is a complex number
-- a local coordinate on $SU(2)/U
(1)=CP^1=S^2$.

The corresponding covariant symbols (the $su(2)$ cs averages)  are
\begin{equation}
S^{+}_c=tr PS^{+}=
\frac{z}{1+|z|^2},\quad S^{-}_c=tr PS^{-}=
\frac{\bar z}{1+|z|^2},\quad S^{z}_c=tr PS^{z}=
\frac{1}{2}\frac{1-|z|^2}{1+|z|^2},
\label{1.0}\end{equation}
where the projection
operator $P:=|z\rangle\langle z|.$

A classical image of the AF Heisenberg model
\begin{equation}
H=J\sum_{ij}\vec S_i\vec S_j, \quad J>0,
\label{eq:1.1}\end{equation}
reads
\begin{equation}
H^ c=tr PH = J\sum_{ij}\vec S^c_i\vec S^c_j=J\sum_{ij}|\langle z_i|z_j\rangle|^2=J\sum_{ij}tr (P_iP_j).
\label{eq:1.2}\end{equation}

The Neel ground state implies staggered classical spin moments so that two neighboring-site spins point in the opposite directions:
$\vec S^c_i, i \in A, \,\, -\vec S^c_i, i \in B, A \bigoplus B=L$
\begin{equation}
H^ c_{AF}= -J\sum_{ij}\vec S^c_i\vec S^c_j=-J\sum_{ij}\left(S^c_{zi}S^c_{zj}+\frac{1}{2}(S^+_{ci}S^-_{cj}+S^-_{ci}S^+_{cj})\right).
\label{eq:1.3}\end{equation}
Here the vectors $\vec S^c_i$ are given by Eqs.(\ref{1.0}) in which the $z_i$ do not depend on time (quantum fluctuations
are ignored). Equation (\ref{eq:1.3}) follows upon the change $z_i\to -1/\bar z_i,\, i\in B.$ Under such a change of
variables $P_i\to 1-P_i$ so that
\begin{equation}
H^ c_{AF}= J\sum_{ij}tr\, P_i(1-P_j)= J\sum_{ij}(1-tr\,P_iP_j)=J\sum_{ij}(1-e^{\Gamma_{ij}}),
\label{eq:1.33}\end{equation}
where $\Gamma_{ij}= F_{ij}+F_{ji}-F_{ii}-F_{jj},$ and the $CP^{1}$ Kaehler potential reads
$F_{ij}=2s\log(1+\bar z_iz_j).$
Small quantum fluctuation around the AF classical ground state, can be incorporated
along the lines depicted in the preceding Section to arrive at the $CP^1$ NLSM
given by Eq.(\ref{eq:6.8}) where one sets $\xi_i=\bar\xi_i\equiv 0.$

Let us now turn to the conjugate representation.
All relevant quantities are
marked by the "tilde" sign.
The irrep conjugated to the fundamental one is generated by the operators
\begin{equation}
\tilde S^+=-S^-,\quad \tilde S^-=-S^+,\quad \tilde S_z=-S_z.
\label{conj}\end{equation}
One therefore gets
\begin{equation}
H^ c_{AF}\to \tilde H^ c_{AF}=-J\sum_{ij}\left(S^c_{zi}S^c_{zj}+\frac{1}{2}(S^+_{ci}S^+_{cj}+S^-_{ci}S^-_{cj})\right).
\label{eq:1.4}\end{equation}
Upon the operation of complex conjugation, $z\to \bar z$, this reduces back to Eq.(\ref{eq:1.3}).

In the irrep conjugated to the fundamental
one we get
$$\tilde {|z\rangle} =(1+|z|^2)^{-1/2}e^{-\bar zS^{+}}|-\rangle=(1+|z|^2)^{-1/2}
(|-\rangle-\bar z|+\rangle).$$
This yields
\begin{equation}
\tilde P(\bar z,z)=\tilde{|z\rangle}\tilde{\langle z|}=1-P(\bar z,z)
\label{eq:1.5}\end{equation}

The $su(2)$ case is special in that any $su(2)$ irrep is equivalent to its
conjugate representation.  Therefore there must be an operator
$U_g,\,g\in SU(2)$
such that $(U_g^{\dagger}PU_g)(\bar z,z)=P(\bar{gz},gz)= \tilde P^*(\bar z,z)$. One
can easily find a required $SU(2)$ action on $CP^1$ in the form $g(z)=-1/z$.
As a result, $P^*(\overline{g(z)},g(z))=\tilde P(\bar z,z),$ and we get back to Eq.(\ref{eq:1.33}).
Note that $tr P_i=tr (1-P_i)=1$ which indicates that all the states on sublattices $A$ and $B$ can be
properly normalized. To summarize, we see that in this case the Neel state can be arranged on the neighboring sites
in a self-consistent manner.

\section{Appendix B}

Let us briefly comment on the $SU(2|1)$ lattice action in the conjugate irrep.

In the $su(2|1)$ fundamental irrep, conjugate to the $(q,q)$ one, we get 
$$ \tilde{ Q^{+}}=-Q^{-},\, \tilde{Q^{-}}=-Q^{+},\, \tilde Q_3=-Q_3,\,
\tilde B=-B,$$
$$ \tilde V_-=-W_+,\, \tilde V_+=W_-,\,\tilde W_+=V_-,\, \tilde W_-=-V_+,$$
where the "tilde" sign indicates the conjugate irrep. It can be checked that the "tilde" operators fulfil the $su(2|1)$ commutation relations. 

In this representation, the $su(2|1)$ coherent state takes the form
\begin{equation}
\tilde{\langle z,\xi|}={\cal N} \langle -1/2,1/2,-1/2|\, e^{-\bar\xi W_+-\bar zQ_+},
\label{b1}\end{equation}
which yields 
\begin{equation}
\tilde{|z,\xi\rangle}=(1+|z|^2 +\bar \xi\xi)^{-1/2}
(|-1/2,1/2,-1/2\rangle -z |-1/2,1/2,1/2\rangle +\xi |-1, 0,0\rangle),
\label{b2}\end{equation}

The classical images of the $su(2|1)$ generators are then found to be:

$$(\tilde{Q^{+}})^c=-(Q^{-})^c,\, (\tilde{Q^{-}})^c=-(Q^{+})^c,\, (\tilde {Q_3})^c=-(Q_3)^c,\,
(\tilde B)^c=-B^c,$$
$$ (\tilde {V_-})^c=-W_+^c,\, (\tilde {V_+})^c=W^c_-,\,(\tilde {W_+})^c=V^c_-,\, 
(\tilde{ W_-})^c=-V_+^c,$$
where $\tilde{ A^c}=\tilde{\langle z,\xi|}\tilde A\tilde{|z,\xi\rangle}$ and the quantities $A^c$ are given by Eqs.(\ref{eq:2.2}).

The lattice Euclidean action becomes
$$S=i\sum_i\oint a_i-\int H_{SUSY}^{cl}(t)dt,$$
where we now have  
\begin{equation}
H^{cl}_{SUSY}= t\sum_{<ij>}str (Q_i\tilde {Q_j}).
\label{eq:b3}\end{equation}
It can easily be seen that $\tilde {Q_j}\neq -Q_j$. Besides, $ \tilde {a_j}$ --the Berry phase in the conjugate representation-- coincides with $a_j$.
As a result, our approach developed in Section IV cannot be applied in this case. 
In view of this observation, it remains unclear whether the SUSY superspin model based on the alternating group representations and the one  based on the generalized AF Neel 
superspin representation lead, in the low-energy limit, to
formally identical actions. This is presumably true in the large-$q$ limit, however, a full comparison of the microscopic derivation of the both models 
is needed to draw a precise conclusion.


\begin{thebibliography}{99}
\bibitem{3d}
For the $3d\,\,t-J$ continuum model,  a relevant mapping is
$\varphi:\,(x_1,x_2,t)=S^3\to CP^{1|1}.$ A natural choice is then
$\Omega^{(4)}=da\wedge da$ --
the second Chern character --
its Chern-Simons form serving as a Wess-Zumino term, $ S_{WZ}=\int_{CP^{1|1}}
a\wedge da.$
Being restricted to the Hopf map $S^3\to CP^1$, which
corresponds to the continuum limit for a $2d$ quantum AF, it
vanishes in accordance with the direct large-spin limit calculations.
\end{thebibliography}
\end{document}